\documentclass[%
 aip,
 amsmath,amssymb,
 reprint,%
]{revtex4-1}

\usepackage{graphicx}
\usepackage{dcolumn}
\usepackage{bm}

\usepackage[utf8]{inputenc}
\usepackage[T1]{fontenc}
\usepackage{mathptmx}

\begin{document}

\preprint{AIP/123-QED}

\title{Influences of Dielectric Constant and Scan Rate on Hysteresis Effect in Perovskite Solar Cell with Simulation and Experimental Analyses}

\author{Jun-Yu Huang}
\affiliation{Graduate Institute of Photonics and Optoelectronics and Department of Electrical Engineering, National Taiwan University, Taipei 10617, Taiwan}
\author{You-Wei Yang}
\affiliation{Department of Electro-Optical Engineering, National Taipei University of Technology, Taipei 10608, Taiwan}
\author{Wei-Hsuan Hsu}
\affiliation{Department of Electro-Optical Engineering, National Taipei University of Technology, Taipei 10608, Taiwan}
\author{En-Wen Chang}%
\affiliation{Graduate Institute of Photonics and Optoelectronics and Department of Electrical Engineering, National Taiwan University, Taipei 10617, Taiwan}
\author{Mei-Hsin Chen}
\affiliation{Department of Electro-Optical Engineering, National Taipei University of Technology, Taipei 10608, Taiwan}
\author{Yuh-Renn Wu}
\homepage{http://yrwu-wk.ee.ntu.edu.tw/}
\email{yrwu@ntu.edu.tw}
\affiliation{Graduate Institute of Photonics and Optoelectronics and Department of Electrical Engineering, National Taiwan University, Taipei 10617, Taiwan}
\date{\today}

\begin{abstract}
In this work, perovskite solar cells (PSCs) with different transport layers were fabricated to understand the hysteresis phenomenon under a series of scan rates. The experimental results show that hysteresis phenomenon would be affected by the dielectric constant of transport layers and scan rate significantly. Also, a modified Poisson and drift-diffusion solver coupled with a fully time-dependent ion migration model is developed and applied to demonstrate current--voltage hysteresis behavior and analyze how ion migration affects the performance and hysteresis of PSCs. In addition, this model was optimized for carrier transportation of organic materials, which can simulate organic transport layer correctly to discuss the change in materials of transport layer without heavily doping in organic transport layer. The modeling results reveal the physical mechanism of hysteresis induced by ion migration. The most crucial factor in the hysteresis behavior is the built-in electric field of the perovskite. Using the modified solver, the non-linear hysteresis curves are demonstrated under different scan rates, and the mechanism of the hysteresis behavior is explained. The findings reveal why the change in hysteresis degree with scan rate is Gaussian shaped rather than monotonic. Additionally, some factors affecting the degree of hysteresis are simulated and discussed. The main factors contributing to the degree of hysteresis are determined to be the degree of degradation in the perovskite material, the quality of the perovskite crystal, and the materials that make up the transport layer, which corresponds to the total ion density, carrier lifetime of perovskite, and the properties of the transport layer, respectively. Finally, we reveal that the dielectric constant of the transport layer is a key factor affecting hysteresis in perovskite solar cells; a lower dielectric constant corresponds to a higher electric field of the transport layer. Hence, if the electric field of the perovskite material is small, the degree of hysteresis is small and vice versa.
\end{abstract}

\maketitle

\section{Introduction}
Since the first metal halide perovskite solar cell (PSC) was fabricated by Prof. Miyasaka \cite{kojima2009organometal}, many researchers have endeavored to improve the performance of PSCs, which exhibit interesting optical and electrical properties. As a result, the PSC power conversion efficiency has been enhanced to 25\% for single-junction devices and 28\% for tandem-structured devices in recent years \cite{NREL}. PSCs are now regarded as one of the most promising candidates for the next generation of solar cells.  

However, the instability of PSC devices remains a challenging problem; PSC devices hydrolyze within few hours when exposed to the air. Hence, encapsulation technology is quite important for the commercialization of PSC devices \cite{zhao2016investigation,wang2016stability}. Another stability issue in PSC devices is the hysteresis of the current--voltage ($J$--$V$) characteristics, which is a common phenomenon observed in the measurement of $J$--$V$ curves. Hysteresis is a complicated problem that may be affected by the scan direction, scan rate, measurement conditions, structure of the PSC device, and other factors. Due to the hysteresis phenomenon, reproducing and correctly measuring the performances of PSC devices is difficult. The hysteresis of the $J$--$V$ characteristics was first reported by Prof. Snaith \cite{snaith2014anomalous}. Although the origin of the hysteresis phenomenon remains debated \cite{schulz2019halide,liu2019fundamental}, three likely causes have been identified. First, some reports have indicated that the ferroelectric polarization causes the hysteresis at the interface of the perovskite material \cite{ferro1,ferro2,ferro3,ferro4}. However, direct evidence to support this behavior in perovskite is lacking. The second proposed explanation is interfacial defect and charge accumulation. Typical PSC structures are planar n-i-p (TiO$\rm _2$/perovskite/Spiro-OMeTAD) and p-i-n (PEDOT:PSS/perovskite/PCBM) structures \cite{ke2014perovskite,long2015dopants}. However, previous studies showed that hysteresis mainly occurs in the n-i-p structure because TiO$\rm _2$ is a mesoscopic layer, and a defect state exists at the TiO$\rm _2$/perovskite interface \cite{li2017extrinsic,okada2019perovskite,etgar2012mesoscopic}. Therefore, hysteresis is more serious in the n-i-p structure due to trapping and de-trapping \cite{neukom2017perovskite}. Finally, hysteresis has been attributed to ion migration in perovskite. Both direct and indirect evidences of ion migration in perovskite have been reported in previous studies \cite{yuan2016ion,calado2016evidence,xing2016ultrafast,lee2017direct,zhang2019phase}. Hence, ion migration might be the primary factor responsible for the hysteresis of the $J$--$V$ characteristics. This effect can be explained as follows. The ions are movable, and their distribution is mainly determined by the built-in electric field. Ions accumulate at the hole transport layer (HTL)/perovskite and the electron transport layer (ETL)/perovskite interfaces. Hence, different current densities are obtained when bias is applied at the same voltage.

Recent modeling studies on PSC devices have mainly focused on (1) the mechanism of ion accumulation, (2) how ion accumulation affects hysteresis, and (3) the effects of different factors of perovskite on hysteresis \cite{richardson2016can,ravishankar2017surface,huang2020detrimental,jacobs2017hysteresis,xiang2019effect,richardson2016can}. To simplify the model, some assumptions have been made about the transport layer of carrier injection with heavily acceptor/donor doping in previous works \cite{bertoluzzi2020mobile,chen2020correlating,walter2018transient,jacobs2018two,van2015modeling}. However, existing doping techniques cannot be applied in the organic transport layer, and some studies have shown that the organic material is an essential factor in the hysteresis behavior \cite{fang2017functions}. To demonstrate the characteristics of carrier transport in organic materials, we applied a Gaussian density of state in a Poisson and drift-diffusion (Poisson-DD) solver recently \cite{huang2019optimization,huang2020analysis,huang2020revealing}. 

Previous studies have revealed three possible explanations for the different hysteresis behaviors of the n-i-p and p-i-n structures: (1) different ETLs and HTLs (organic material and metallic oxide) \cite{kang2019current,ayguler2018influence,sun2020impact,courtier2019transport}; (2) different degrees of degradation; and (3) different perovskite quality \cite{heo2016highly,lee2019verification}. Moreover, many studies have reported that hysteresis is affected by the scan rate of the voltage measurement in modeling and experiments \cite{HFSR1,HFSR2,HFSR3}. Hence, in this study, both experiments and modeling were presented. In the experiment's part, a series of devices with different voltage scan rates and different materials of transport layer were fabricated to understand the performances of hysteresis effect under different conditions. For example, replacing TiO$\rm _2$ as the ETL material with SnO$\rm _2$ can reduce hysteresis in the n-i-p structure. The dielectric constants of SnO$\rm _2$, and TiO$\rm _2$ are 9, and 31, respectively. Thus, the change in hysteresis behavior might be related to the dielectric constant of the transport layer \cite{courtier2019transport}. In the modeling's part, some studies used heavily doping in the transport layer to model the behavior of organic material, which can discuss the trend in hysteresis with different transport layers \cite{courtier2019transport,jacobs2018two}. However, it is hard to quantize the result and discuss the mechanism in the real case. Because the change in hysteresis with different transport layers can be related to conduction and valence band bending, but these effects would be screen by heavily doping in transport layers, which leads to explaining the mechanism is difficult. Hence, this program utilized Gaussian density of states to simulate correctly different effects of transport layer including organic material, metal oxides and perovskite instead of heavily doping in transport layer \cite{bassler1993charge}. And a fully time-dependent ion migration model is coupled with this modified Poisson-DD solver. Using these methods, our modeling results of hysteresis curves are similar to our experimental results and can be used to analyze and explain the mechanism of the scan rate-dependent hysteresis of $J$--$V$ characteristics.

\section{Methodology} 
In this work, both experiments and modeling are established to understand the mechanism of hysteresis clearly. The experiment section includes the protocols and details of fabrication for each layer and the measurement method for device characteristics. The simulations can be divided into two parts: optical and electrical properties. The finite-difference time-domain (FD-TD) method was utilized to model optical properties, while the time-dependent ion migration model coupled with the Poisson-DD solver was used for the modeling of electrical properties. A Gaussian density of state was implemented to describe the properties of organic materials. The modeling structure is shown in figure \ref{Fig1}(a), and the parameters used in this work are shown in Table \ref{tab1}.

\subsection{Experiment Section}  

\subsubsection{Preparation of Perovskite Solar Cell Device}
The perovskite solar cell devices were fabricated with the following procedures. ITO glass substrates were sequentially cleaned with deionized water, acetone, isopropyl alcohol under an ultrasonic cleaner for 15 minutes, then putting the substrate in UV-Ozone exposed for 15 minutes after drying the substrate. For the SnO$\rm_2$ layer, the SnO$\rm_2$ colloidal solution was spin-coated on ITO substrates at 3000 rpm for 30 seconds, then annealed in air at 180 $\rm^{\circ}$C for 45 minutes; For the TiO$\rm_2$ layer, the precursor solution of TiO$\rm_2$ was spin-coated on ITO substrates at 2000 rpm for 40 seconds, then annealed in air at 150 $\rm^{\circ}$C for 10 minutes. The precursor solution of perovskite was prepared in an N$\rm_2$-filled glovebox by dissolving 578 mg Lead(II) iodide (PbI$\rm_2$) and 200 mg methylammonium iodide (MAI) in 1 ml dimethylformamide (DMF). The precursor of perovskite was spin-coated at 4000 rpm for 20 seconds, 200 $\rm \mu$L of chlorobenzene (CB) was dropped on the film for 7 seconds at the beginning of the program, then forming a uniform thin film. The films were annealed at 100 $\rm^{\circ}$C for 10 minutes in an N$\rm_2$-filled glove box. Spiro-OMeTAD layer was deposited onto the perovskite film by spin-coating the solution of Spiro-OMeTAD with 2000 rpm for 30 seconds. Finally, the Au electrodes (thickness 800 nm) were thermally deposited on the Spiro-OMeTAD electron transporting layer. The active perovskite solar cell area of 0.08 cm$\rm^2$ was defined by the metal mask used during the Au deposition. See the supplementary material for details of device fabrication for each layer (Section Details of Experiments).

\subsubsection{Device Characterization}
The device performances were measured using a source measurement unit instrument (Model 2420 Source Meter, Keithley Instruments) and a solar simulator (Sun 2000 Class A, ABET technologies) under the standard AM 1.5G (100 mW/cm$\rm_2$). The measurement method of cases with different scan rates adjusts the measurement voltage interval to change the scanning speed. In this work, the measurement voltage range is from +1.5 V to -0.5 V for all cases, and the range of scanning rate is from 10 mV/s to 1200 mV/s. The number of measurement data points will decrease, leading to the $J-V$ curve's characteristic not being clear if the scanning rate is more significant than 1200 mV/s.

\subsection{Modeling Section}  

\subsubsection{Simulation of Optical Properties}   
  To model the light absorption in the device, the FD-TD method was implemented to model wave propagation and absorption. The FD-TD program was developed by our lab. The first FD-TD algorithm, which was proposed by Prof. Yee, is based on solving the two major Maxwell equations in the time and space domains, as shown in Eqs. (\ref{eq1}) and (\ref{eq2}) \cite{Yee}:

\begin{equation}
\nabla\times\vec{E}=-\mu \frac{\partial \vec{H}}{\partial t},
\label{eq1}
\end{equation}
and
\begin{equation}
\nabla\times\vec{H}=\varepsilon \frac{\partial \vec{E}}{\partial t},
\label{eq2}
\end{equation}
where $\vec{E}$ is the electric field, $\vec{H}$ is the magnetic field, $\mu$ is the permeability, and $\epsilon$ is the permittivity. 

The details of the FD-TD modeling, including the simulation domain and parameters, and the modeling results are summarized in the Supporting Information (Section FD-TD modeling).

\begin{figure}[!bpht]
\renewcommand{\figurename}{Figure}
\centering
\resizebox{85mm}{!}{\includegraphics{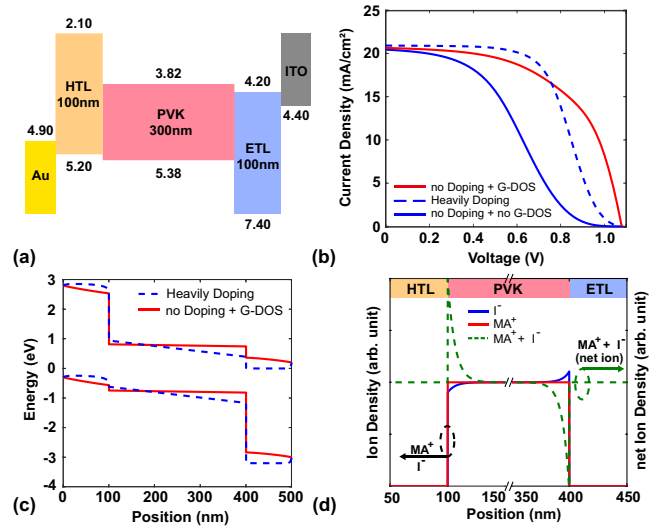}}
\caption{(a) Schematic band diagram of the n-i-p PSC device. (b) The current density for cases of 1. without doping and with Gaussian density of state (G-DOS) 2. heavily doping 3. with G-DOS (c) The conduction and valence band for cases of heavily doping and Gaussian density of state. (d) Schematic showing I$^{-}$, MA$^{+}$, and net ion density.}
\label{Fig1}
\end{figure} 

\subsubsection{Poisson and Drift-Diffusion Solver}

After the optical properties were obtained using the FD-TD solver, the Poisson-DD solver was applied to calculate the electrical characteristics of the device \cite{tsai2020application,chen2018three}. Briefly, this solver is based on a self-consistent solution of the Poisson equation:
\begin{equation}
\nabla_{r}\cdot(\varepsilon\nabla_{r}V(r))=q(n-p+N_{anion}-N_{cation}),
\label{eq3}
\end{equation}
the drift-diffusion equations:
\begin{equation}
J_{n}=-q\mu_{n}n(r)\nabla_{r}V(r)+qD_{n}\nabla_{r}n(r)
\label{eq4}
\end{equation}
and
\begin{equation}
J_{p}=-q\mu_{p}p(r)\nabla_{r}V(r)-qD_{p}\nabla_{r}p(r),
\label{eq5}
\end{equation}
and the continuity equation:
\begin{equation}
\nabla_{r}(J_{n,p})=(R-G_{opt}),
\label{eq6}
\end{equation}
where $q$ is the elementary charge; $\varepsilon$ is the dielectric constant; $V$ is the potential of the device; $n$ and $p$ are the electron density and hole density, respectively; and $N_{cation}$ and $N_{anion}$ are the cation and anion density, which are calculated by the ion migration model.

The drift-diffusion equations are shown in Eqs. (\ref{eq4}) and (\ref{eq5}), where $J$ is the current density, $\mu$ is the mobility, and $ D$ is the diffusion coefficient. The subscripts $e$ and $h$ indicate electrons and holes, respectively. The continuity equation is shown in Eq. (\ref{eq6}), where $R$ is the recombination rate, which is determined by Shockley--Read--Hall ($SRH$) non-radiative recombination, and the radiative recombination coefficient ($B$) is shown in Eqs. (\ref{eq7}) and (\ref{eq8}):

\begin{equation}
R=SRH+Bnp,
\label{eq7}
\end{equation}
\begin{equation}
SRH=\frac{pn-n^{2}_{i}}{\tau_{n}(p+n_{i}exp(\frac{E_{t}}{k_{b}T}))+\tau_{p}(p+p_{i}exp(\frac{E_{t}}{k_{b}T}))},
\label{eq8}
\end{equation}
where $\tau_n$ and $\tau_p$ are the non-radiative carrier lifetimes of electron and hole, respectively; $n_{i}$ is the intrinsic carrier density; and $G_{opt}$ is the generation rate, which is obtained from Eq. (\ref{eq9}):
\begin{equation}
G_{opt}=\frac{1}{\hbar\omega}Re(\nabla\cdot \vec{P})=\frac{nk}{\hbar}\varepsilon|\vec{E}|^{2},
\label{eq9}
\end{equation}
where $\hbar\omega$ is the photon energy; $\vec{P}$ is the Poynting vector; $n$ is the refractive index; $k$ is the extinction coefficient; and $\vec{E}$ is the electric field.

\subsubsection{Gaussian Density of State}
The band edges of organic material are called the highest occupied molecular orbital (HOMO) and lowest unoccupied molecular orbital (LUMO); these are similar to the valance band and conduction band. However, different from semiconductors, the band edges are not abrupt. Hence, some tail states exist between the LUMO and HOMO, meaning that the density of state is a disordered distribution. B$\rm \ddot{a}$ssler et al. proposed that the disordered density of state in an organic material has a Gaussian-like shape \cite{bassler1993charge}. As we had mentioned before, using Gaussian density of state can simulate the transport layer of organic material more correctly, which shows in figures \ref{Fig1}(b)-(c). Figure \ref{Fig1}(b) shows that carriers can not be extracted efficiently as Gaussian density of state is unimplemented. Using heavily doping in transport layers is a common and feasible method to solve this problem, but it can not present the real case. Figure \ref{Fig1}(b) shows using Gaussian density of state can achieve a similar result which implementing heavily doping, and the advantage of this method can be observed in figure \ref{Fig1}(c). The conduction and valence band bending can be demonstrated by using Gaussian density of state, which is a key factor to discuss the mechanism of hysteresis. Hence, Gaussian density of state [Eq. (\ref{eq10})] is implemented to describe the tail states in this work:
\begin{equation}
N_{tail}(E)=N_{t}\frac{1}{\sigma\sqrt{2\pi}}exp\left[-\frac{(E-E_t)^2}{2\sigma^{2}}\right],
\label{eq10}
\end{equation}
where $N_t$ is the total density of state; $\sigma$ is the broadening factor of the Gaussian shape; and $E_{t}$ is the difference between the center of the Gaussian density of state and the LUMO or HOMO. And the Gaussian density of state that have successfully utilized in our previous works \cite{huang2020analysis,huang2019optimization}. Details of the Gaussian density of state, including a schematic and the parameters used in this work, are shown in the Supporting Information (Section Gaussian Density of State). When a Gaussian density of state is applied, the carrier density can be expressed as 
\begin{equation}
n, p_{free}=\int_{-\infty}^{\infty} N_{tail}(E)f_{e,h}(E) \, dE;
\label{eq11}
\end{equation}
otherwise, the carrier density is expressed as
\begin{equation}
n_{free}=\int_{-\infty}^{\infty} \frac{1}{2\pi^2} \Big(\frac{2m^*_{e}}{\hbar^2}\Big)^{\frac{3}{2}} \sqrt{E-E_c}~f_{e}(E) \, dE,
\label{eq12}
\end{equation}
and

\begin{equation}
p_{free}=\int_{-\infty}^{\infty} \frac{1}{2\pi^2} \Big(\frac{2m^*_{h}}{\hbar^2}\Big)^{\frac{3}{2}} \sqrt{E_v-E}~f_{h}(E) \, dE,
\label{eq13}
\end{equation}
where $m_{e,h}$ is the effective mass of the electron or hole, respectively; $E_{c,v}$ is the conduction or valance band, respectively; and $f_{e,h}$ is the Fermi--Dirac distribution function for the electron or hole, respectively.

\subsubsection{Time-Dependent Ion Migration Model}

To evaluate the effect of the voltage scan rate on hysteresis, the time-dependent ion migration model was coupled with the Poisson-DD solver. The ion migration model considers both the anion (iodide, I$^-$) and cation (methyl ammonium, MA$^+$). However, some studies have shown that the mobility of I$^-$ is 10$^{4}$ times faster than the mobility of MA$^+$ \cite{lee2019effect,gautam2020reversible}. Due to the low mobility of MA$^+$, MA$^+$ cannot respond to typical scan rates, which range from 10 to 10$^4$ mV/s. Hence, MA$^+$ is almost immobile in this model. Therefore, although both types of ions are considered in the model, the net ion density in this work mainly reflects the I$^-$ density, which is shown in figure \ref{Fig1}(d) and the net ion density is zero. To evaluate cation and anion migration, the following cation and anion drift-diffusion equations were used:

\begin{equation}
\frac{dN_{cation}}{dt}=\nabla(q\mu_{cation}N_{cation}E-qD_{cation}\nabla N_{cation}),
\label{eq14}
\end{equation}
and
\begin{equation}
\frac{dN_{anion}}{dt}=\nabla(q\mu_{anion}N_{anion}E+qD_{anion}\nabla N_{anion}),
\label{eq15}
\end{equation} 
where $N$ is the ion density distribution; $\mu$ is the ion mobility; $E$ is the electric field; and $D$ is the diffusion coefficient, which is determined by the Einstein relation $D$ = $\mu k_{B}T$, where $k_{B}$ is the Boltzmann constant, and $T$ is room temperature. The subscripts cation and anion indicate MA$^{+}$ and I$^{-}$, respectively. 

\begin{table}[ht]
\centering
\footnotesize
\caption{Parameters used in the modeling in this study}
\begin{tabular}{c c c c c c c c c c c}
\hline\hline
 & HTL & Perovskite & ETL \\
Material & Spiro-OMeTAD & MAPbI$_3$ & TiO$_2$ / SnO$_2$ \\
\hline 
Thickness (nm) & 100 & 300 & 100 \\
Electron effective mass (m$_0$) & -- & 0.25 \cite{aa} & 0.20 \cite{zhang2014new} \\

Hole effective mass (m$_0$) & -- & 0.25 \cite{aa} & 0.25 \cite{zhang2014new} \\

Electron mobility (cm$^{2}$/Vs) & 4$\times 10^{-07}$ \cite{li2017characterization} & 2.0 \cite{maynard2016electron,lim2019elucidating,rolland2018computational} & 1$\times 10^{-03}$ \cite{forro1994high} \\

Hole mobility (cm$^{2}$/Vs) & 3$\times 10^{-03}$ \cite{li2017characterization} & 2.0 \cite{maynard2016electron,lim2019elucidating,rolland2018computational} & 5$\times 10^{-11}$ \cite{forro1994high} \\

$\tau_{n}~/~\tau_{p}$ (ns) & 20 / 20 & 70 / 70 \cite{de2015impact,hoang2020impact,guo2019enhanced} & 350 / 20 \cite{yamada2012determination} \\

Relative permittivity ($\epsilon_r$) & 3 & 25.7 & 31 / 9 \\

Anion mobility (cm$^2$/Vs) & -- &  4$\times 10^{-11}$ \cite{lee2019effect} & -- \\
Cation mobility (cm$^2$/Vs) & -- &  1$\times 10^{-15}$ \cite{lee2019effect} & -- \\ 
 \hline\hline

\label{tab1}
\end{tabular}
\end{table}
\section{Results and Discussion} 
\subsection{Hysteresis of the $J$--$V$ curve based on Ion Accumulation} 
\begin{figure}[!bpht]
\renewcommand{\figurename}{Figure}
\centering
\resizebox{85mm}{!}{\includegraphics{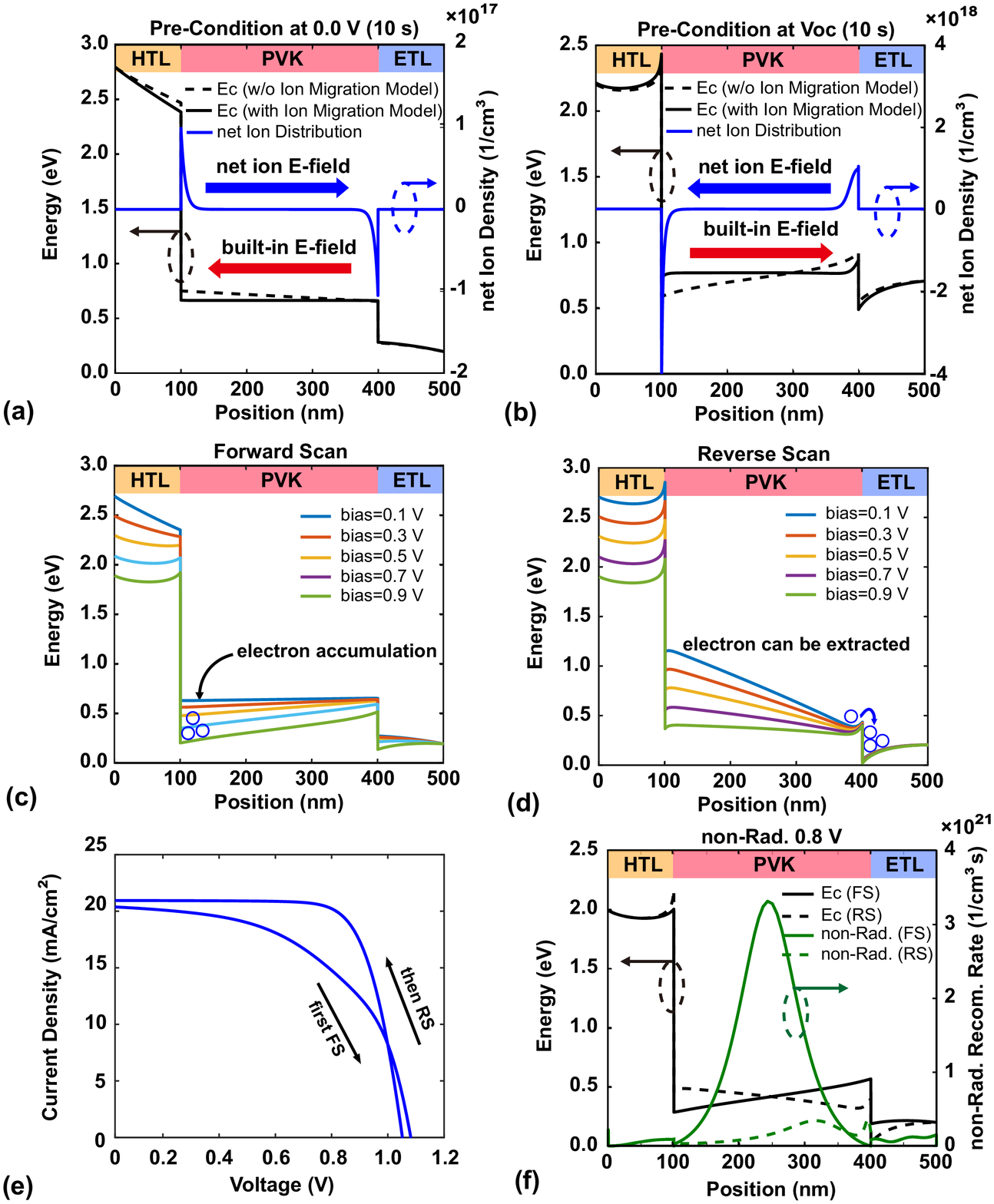}}
\caption{(a) and (b) are band diagram and net ion distribution for preconditioning at 0 V and $\rm V_{oc}$, respectively. (c)--(d) Changes in the band diagram when bias is applied under forward and reverse scanning conditions, respectively. (e) $J$--$V$ characteristics under forward and reverse scanning conditions. (f) Band diagram and non-radiative recombination rate under forward and reverse scanning conditions at a bias of 0.6 V.}
\label{Fig2}
\end{figure}

In the following, the modeling result is utilized to analyze the mechanism of hysteresis phenomenon. To evaluate the effect of ion accumulation, the default scan rate was set to 10$\rm ^{5}~mV/s$. The complete modeling sequence involved the following steps: (1) pre-condition at 0 V for 10 seconds to demonstrate ion accumulation at the interface; (2) calculate the $J$--$V$ curve under forward scanning conditions from 0 V to $\rm V_{oc}$ at a scan rate of 10$\rm ^{5}~mV/s$ (to demonstrate the situation when the ion accumulates at the interface, a faster scan rate is needed in this case); (3) pre-condition at $\rm V_{oc}$ for 10 seconds to evaluate ion accumulation at the interface; and (4) calculate the $J$--$V$ curve under reverse scanning conditions at a scan rate of 10$\rm ^{5}~mV/s$. Figures \ref{Fig2}(a) and \ref{Fig2}(b) show the net ion distribution and the conduction band after pre-conditioning at 0 V and $\rm V_{oc}$ for 10 seconds, respectively. To achieve a steady-state, the ions redistribute to screen the built-in electric field, generating an ion-induced electric field with a direction opposite the built-in electric field. Figures \ref{Fig2}(a) and \ref{Fig2}(b) show that the built-in electric field is screened by the ion-induced electric field; thus, ion accumulation at the interface can be observed. Because the direction of the built-in electric field is opposite that for the pre-condition at 0 V and $\rm V_{oc}$, the net ion distribution is also opposite. Bias was then applied from 0 V to $\rm V_{oc}$ (forward scanning) and from $\rm V_{oc}$ to 0 V (reverse scanning). Figure \ref{Fig2}(e) shows the $J$--$V$ characteristics under the forward and reverse scanning conditions, and apparent hysteresis behavior can be observed. The hysteresis phenomenon can be explained by the conduction bands in the two cases. Figures \ref{Fig2}(c) and \ref{Fig2}(d) show the conduction bands for the forward and reverse scanning, respectively. For the forward scanning, the direction of the ion-induced electric field is from the HTL to the ETL, the same direction as the applied electric field. Figure \ref{Fig2}(c) shows that electrons are accumulated at the side of the HTL, leading to a decrease in carrier extraction. It is not an ideal condition for carrier extraction when the direction of the ion-induced electric field is the same as that of the applied electric field. For the reverse scanning, the direction of the ion-induced electric field is from the ETL to the HTL, the opposite direction as the applied electric field. Under this condition, carriers can be easily extracted because the applied electric field is screened by the ion-induced electric field. Unlike the forward scanning, the ion-induced electric field can improve carrier extraction in this situation. The difference in the non-radiative recombination rate distribution between the forward and reverse scanning is shown in figure \ref{Fig2}(f). Hence, the hysteresis phenomenon can be attributed to the difference in non-radiative recombination between the forward and reverse scanning, which is caused by the ion-induced electric field. 

In summary, interfacial ion accumulation is caused by the built-in electric field and leads to an opposite ion-induced electric field in the active layer. Because carrier extraction is affected by the applied electric field, the ion-induced electric field has the opposite direction under the forward and reverse scanning. Therefore, carrier extraction differs between these two cases, leading to different non-radiative recombination at the same voltage. This results in different current densities when measured at the same voltage. Hence, the hysteresis phenomenon is observed. In the next section, the effect of the voltage scan rate is discussed.

\subsection{Effect of Scan Rate on Hysteresis}
\begin{figure}[!bpht]
\renewcommand{\figurename}{Figure}
\centering
\resizebox{85mm}{!}{\includegraphics{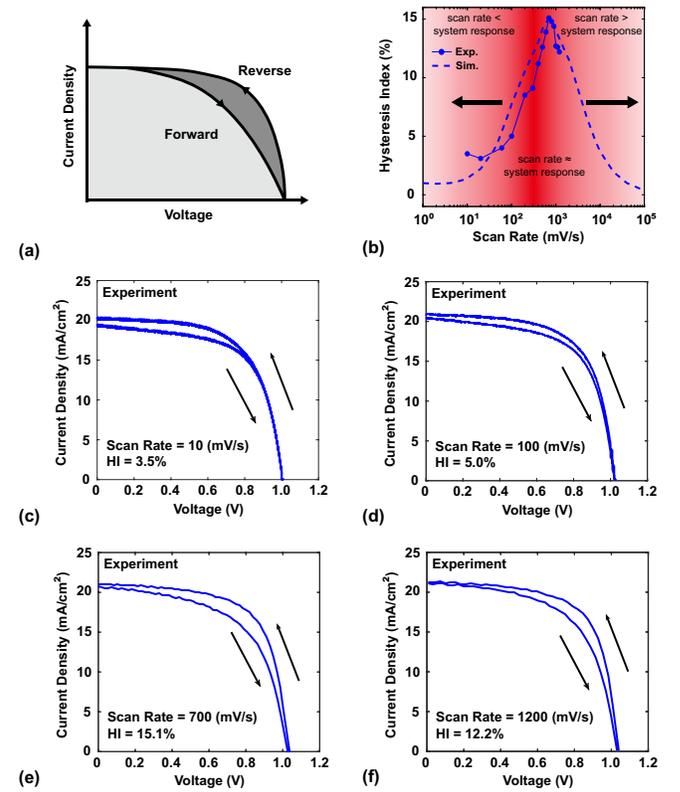}}
\caption{(a) Schematic showing the definition of hysteresis index. (b) Change in hysteresis factor with scan rate, the solid line and dash line are experiment and modeling results, respectively. (c)--(f) $J$--$V$ characteristics in experiments at scan rates of 10, 100, 700, and 1200 mV/s, respectively.}
\label{Fig3}
\end{figure} 

Several studies have shown that the hysteresis in the $J$--$V$ characteristics depends on the scan rate of the voltage measurement \cite{HFSR1,HFSR2,HFSR3}. Tress et al. proposed an explanation for how the scan rate affects hysteresis; they found that hysteresis only occurs within a certain range of scan rates, while it disappears when the scan rate is outside of this range \cite{tress2015}. 

To discuss this issue clearly, in this work, both experiments and modeling are demonstrated, the structures are the same as Au/Spiro-OMeDA/MAPbI$\rm_3$/TiO$\rm_2$/ITO. Before discussing the effect of scan rate, it is necessary to define a factor to quantify the degree of hysteresis. Several such factors have been defined in previous studies \cite{habisreutinger2018hysteresis,courtier2019transport}. In this work, the hysteresis index (HI) is defined as:

\begin{equation}
HI=\frac{PCE_{RS}-PCE_{FS}}{PCE_{RS}}~,
\label{eq16}
\end{equation} 
where $\rm PCE_{RS}$ and $\rm PCE_{RS}$ are power conversion efficiency under the forward and reverse scanning conditions, respectively. Figure \ref{Fig3}(a) schematically shows the definition of HI.

Figure \ref{Fig3}(b) shows that both of the experimental and modeling results show a similar trend. The experiment shows there are different hysteresis degrees under the different scan rates, and the hysteresis pattern like a Gaussian distribution, which means a lower scan rate (below 100 mV/s) shows lower hysteresis, then hysteresis phenomenon would get more serious with scan rate increasing. But hysteresis decrease as the scan rate is higher than 700 mV/s. Figures \ref{Fig3}(c)--\ref{Fig3}(f) show that the degree of hysteresis changes with scan rate (see the Supporting Information Section Experimental Data to find all experimental data in $J$--$V$ curve). In the following section, the simulation program was utilized to explain the mechanism of these results. The modeling results aim to demonstrate the effect of scan rate on hysteresis along with the associated mechanism. 

\begin{figure}[!bpht]
\renewcommand{\figurename}{Figure}
\centering
\resizebox{85mm}{!}{\includegraphics{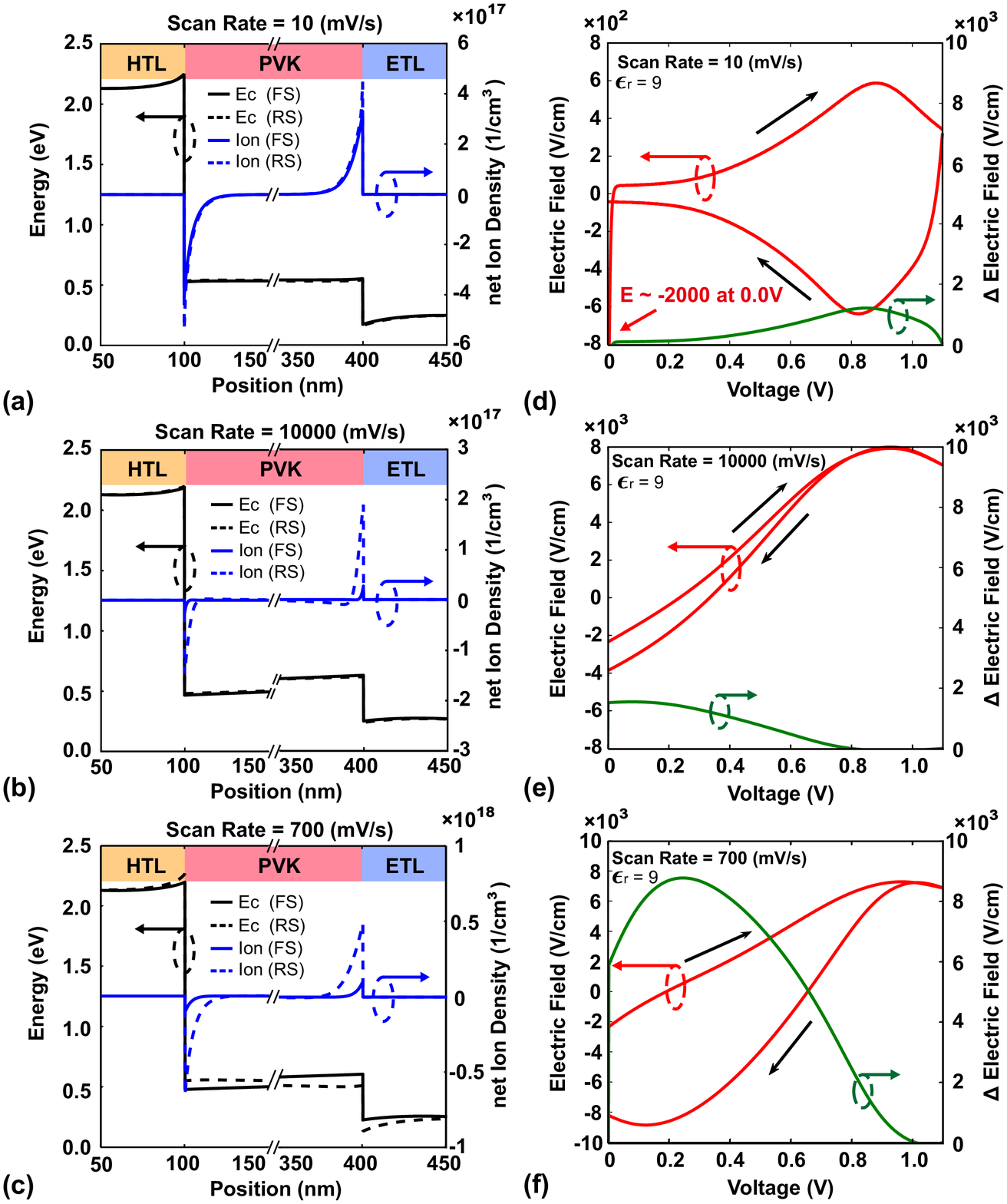}}
\caption{(a)--(c) Conduction bands and net ion density distributions at the voltage of 0.6 V for scan rates of 10, $\rm 10^{4}$, and 700 mV/s, respectively. (d)--(f) Electric field and change in electric field as functions of bias voltage for scan rates of 10, $\rm 10^{4}$, and 700 mV/s, respectively.}
\label{Fig4}
\end{figure}

The protocol for the simulation described in the next section involved the following steps (different from the previous section): (1) calculate the current density from a bias of 0 V to $\rm V_{oc}$ under a specific scan rate; and (2) after forward scanning, immediately calculate the current density from a bias of $\rm V_{oc}$ to 0 V using the same scan rate. The scan rate in this simulation ranged from 1 to 10$^{5}$ mV/s. The modeling results of HI at scanning rates of 1--$\rm 10^5$ mV/s are shown in figure \ref{Fig3}(b). The maximum HI is observed at the scan rate of 700 mV/s, and the degree of hysteresis gradually decreases as the scan rate increases or decreases from this value. As mentioned before, Tress et al. reported this phenomenon in a previous work \cite{tress2015}. Levine et al. also reported similar behavior; they termed the appropriate scan rate as the ``system response region`` and found that hysteresis also occurs in the p-i-n structure but with a different system response region than for the n-i-p structure \cite{levine2016interface}. To understand the mechanism of this effect, three scan rates (10, 700, and 10$\rm ^{4}~mV/s$, corresponding to slow, appropriate, and fast scan rates, respectively) were chosen for further evaluation. The conduction bands, net ion densities, and electric fields for these cases are shown in figures \ref{Fig4}(a)--\ref{Fig4}(f).

Figure \ref{Fig4}(a) shows that under the forward and reverse scanning conditions at the scan rate of 10 mV/s, the net ion densities are similar, and the conduction bands are flat. Because the ion movement responds to the scan rate, the net ion distribution depends on the specific bias when the scan rate is slow. Figure \ref{Fig4}(d) shows that the electric field of the active layer is quite small under both forward and reverse scanning conditions. Because the ions have sufficient time to accumulate at the interface, the built-in electric field is completely screened by the ion-induced electric field. Hence, the $J$--$V$ characteristics are nearly the same under the forward and reverse scanning conditions, and the hysteresis is weak. For the case of the fast scan rate [figure \ref{Fig4}(b)], the velocity of ion redistribution cannot respond to the high scan rate (10$\rm ^{4}~mV/s$), and less ion accumulation is observed compared to the other two scan rates. Therefore, the effect of the screening of the built-in electric field is not apparent. Figure \ref{Fig4}(e) shows that the electric field is almost the same under the forward and reverse scanning conditions, and the hysteresis is weak. Figure \ref{Fig4}(c) shows the net ion density and conduction band distribution under forward and reverse scanning conditions at 0.6 V. The difference in ion density between the forward and reverse scanning is quite large; thus, the conduction band also differs between the forward and reverse scanning. Figure \ref{Fig4}(f) shows that the difference in the electric field (green line) between the forward and reverse scanning at the same voltage is larger at the scan rate of 700~mV/s than at the other two scan rates. Hence, the scan rate of 700~mV/s is the system response scan rate in this case. 

In summary, hysteresis cannot occur when the scan rate is larger or smaller than the system response scan rate. At a scan rate that is lower than the system response rate, the net ion distribution depends on the bias, and the ions have sufficient time to accumulate at the interface. Therefore, the measurement results are the same under the forward and reverse scanning. For scan rates larger than the system response rate, the ions cannot respond to the voltage scan, and the ion effect is thus quite weak. Thus, the measurement results are again the same under the forward and reverse scanning. Animations for the above cases are presented as supporting information on the website, clearly demonstrating the above-mentioned mechanisms. As mentioned earlier, some studies have shown that the system response region is related to the degree of perovskite dissociation, the quality of perovskite, and the properties of the transport layer. These are the possible reasons for the observed difference in HIs between the n-i-p and p-i-n structures. The following section presents the simulation results for different net ion densities, carrier lifetimes, and properties of the transport layer. 


\subsection{Effect of transport layer's Dielectric Constant on Hysteresis}%
As mentioned above, PSCs with n-i-p structures exhibit significant hysteresis, while hysteresis is weak in devices with p-i-n structures. Some studies have indicated that the properties of the transport layer are important in the degree of hysteresis \cite{fang2017functions,kang2019current}. Fang et al. reported that PCBM, a fullerene material, plays an important role in the p-i-n structure. They attempted to remove PCBM and adjust the thickness of PCBM and discovered that using PCBM can reduce the hysteresis due to the trapping of ions by PCBM. However, some studies have shown that hysteresis can be further reduced by replacing TiO$\rm _2$ with SnO$\rm _2$ as the ETL in the n-i-p structure \cite{baena2015highly,lee2019band,tavakoli2018mesoscopic}. Previous studies have attributed the reduced hysteresis to the permittivity of the transport layer \cite{ayguler2018influence,sun2020impact,courtier2019transport}. In this section, both experiments and modeling were developed to support this viewpoint and analyze the mechanism. 

\begin{figure}[!bpht]
\renewcommand{\figurename}{Figure}
\centering
\resizebox{85mm}{!}{\includegraphics{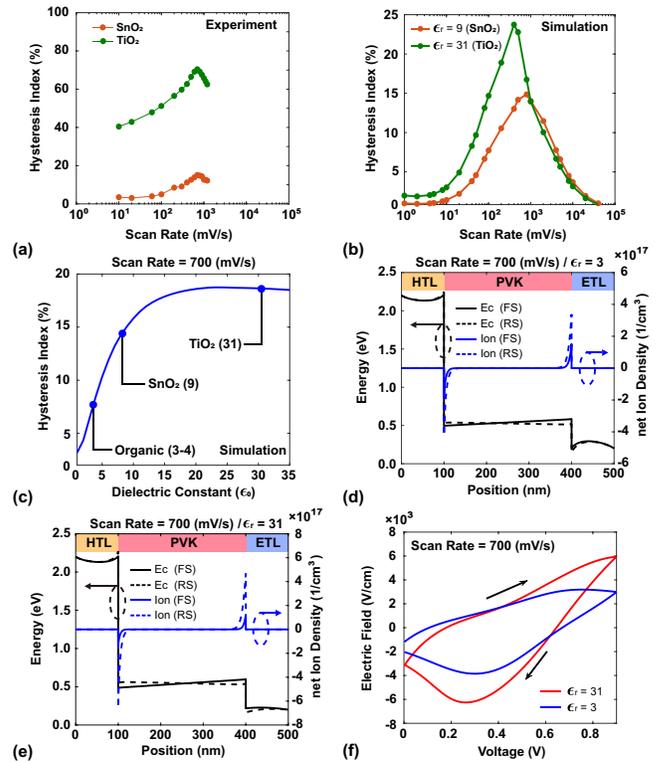}}
\caption{(a) HI values at different scan rates for experiments with ETL of Sn$\rm_2$ and TiO$\rm_2$. (b) HI values at different scan rates for modeling results with the dielectric constants of 9 and 31. (c) HI values for different dielectric constants at the fixed scan rate of 700$ ~\rm mV/s$. (d)--(e) Conduction band and net ion density distribution at the voltage of 0.6 V for dielectric constants of 3 and 31, respectively. (f) Electric field vs. voltage under forward and reverse scanning.}
\label{Fig5}
\end{figure}

Figure \ref{Fig5}(a) also shows this trend. We replaced the materials of ETL from SnO$_{2}$ ($\epsilon_{r}$=9) to TiO$_{2}$ ($\epsilon_{r}$=31) in the experiment, a significant increase of hysteresis index can be observed (see Supporting Information Section Experimental Data for the complete experimental data under different scan rates). The experimental data of TiO$_2$ shows a serious hysteresis. The reason is surface traps at interface between TiO$_2$ and perovskite layer besides a higher dielectric constant \cite{you2018biopolymer}. By investigating and comparing reported cases in which the hysteresis was reduced and the experiments in this work, we can note that the dielectric constant is an important factor. The dielectric constants of PCBM, SnO$\rm _2$, and TiO$\rm _2$ are 3, 9, and 31, respectively. The dielectric constant reflects the degree of polarization in the material; thus, a higher dielectric constant corresponds to a lower electric field and vice versa. The dielectric constants of metallic oxides and organic materials are significantly different. Therefore, the difference in dielectric constant is a possible explanation for the difference in hysteresis degree. In this section, in order to discuss the effects of different dielectric constants purely, the dielectric constant of the ETL was changed from 3 (organic material) to 31 (metallic oxide), while the other parameters remained the same (Table \ref{tab1}).

Figure \ref{Fig5}(b) shows the changes in HI with scan rate for the dielectric constants of 9 and 31 in this simulation. Hysteresis is reduced when the dielectric constant changes from 31 to 9. The scan rate of 700$\rm ~mV/s$ was chosen to analyze the effect of dielectric constant on HI. As shown in figure \ref{Fig5}(b), the HI is lower when the dielectric constant is 3 (corresponding to PCBM as the ETL) than when the dielectric constant is 31 (TiO$\rm _2$ as the ETL). The HI for the dielectric constant of 9 (SnO$\rm _2$ as the ETL) is lower than that for the dielectric constant of 31 (TiO$\rm _2$). The cases with the dielectric constants of 3 and 31 were chosen to explain the observed effect, as shown in figure \ref{Fig5}(c). Figures \ref{Fig5}(d) and \ref{Fig5}(e) show the conduction band and net ion density for the two cases at the voltage of 0.6 V, respectively. For the conduction band, the band bending of the ETL is stronger when the dielectric constant is 3 than when it is 31 [figure \ref{Fig5}(d)]. The lower dielectric constant results in a stronger degree of polarization. Thus, the electric field of the ETL is higher when the dielectric constant is 3 than when it is 31. The entire device can be simplified as a series circuit; hence, the built-in electric field of perovskite increases when the electric field of the ETL decreases. Therefore, more ion accumulation also occurs, as observed in figures \ref{Fig5}(d) and \ref{Fig5}(e). Figure \ref{Fig5}(f) shows the electric field in the middle of the active layer. Compared to when the dielectric constant is 3, the electric field is stronger when the dielectric constant is 31, and the degree of hysteresis is thus higher. 

In conclusion, the ETL plays an important role in the degree of hysteresis. The effect of the ETL can be attributed to various properties of the ETL, especially the dielectric constant. Due to the polarization of the ETL, the electric field of the ETL is weak when the ETL material has a high dielectric constant (e.g., metallic oxide). This causes the built-in electric field of perovskite to be strong, leading to serious hysteresis because the strong built-in electric field results in serious ion accumulation. This explains why studies have found reduced hysteresis when the ETL is changed from TiO$\rm _2$ to another metallic oxide with a lower dielectric constant, such as SnO$\rm _2$.

\subsection{Effect of Total Ion Density and Carrier Lifetime on Hysteresis}
Some studies have found that degradation and quality of perovskite are important factors in hysteresis phenomenon \cite{levine2016interface,heo2016highly,lee2019verification}. However, discussing these two factors is quite hard in experiments. The reason is that degradation and quality of perovskite depend on the type of perovskite, doping other materials (such as K, Rb, and Li), or crystallizing of perovskite \cite{saliba2016incorporation,bu2017novel,kim2017engineering}. Hence, modeling is utilized to discuss these two factors without experiments in this section.

A previous study reported a total ion density of nearly $\rm 10^{18}~cm^{-3}$ in perovskite \cite{jacobs2017hysteresis}. However, recent studies have shown that the total ion density is determined by multiple factors, including the process, material, and structure. As we mentioned before, some materials have been doped into the perovskite layer to capture ions ($\rm I^{-}$), including K, Rb, and Li, thereby reducing hysteresis. In this section, the total ion density ($N_{ion}$) is changed from $\rm 10^{16}$ to $\rm 5\times 10^{18}~cm^{-3}$ while the other parameters remain the same.

\begin{figure}[!bpht]
\renewcommand{\figurename}{Figure}
\centering
\resizebox{85mm}{!}{\includegraphics{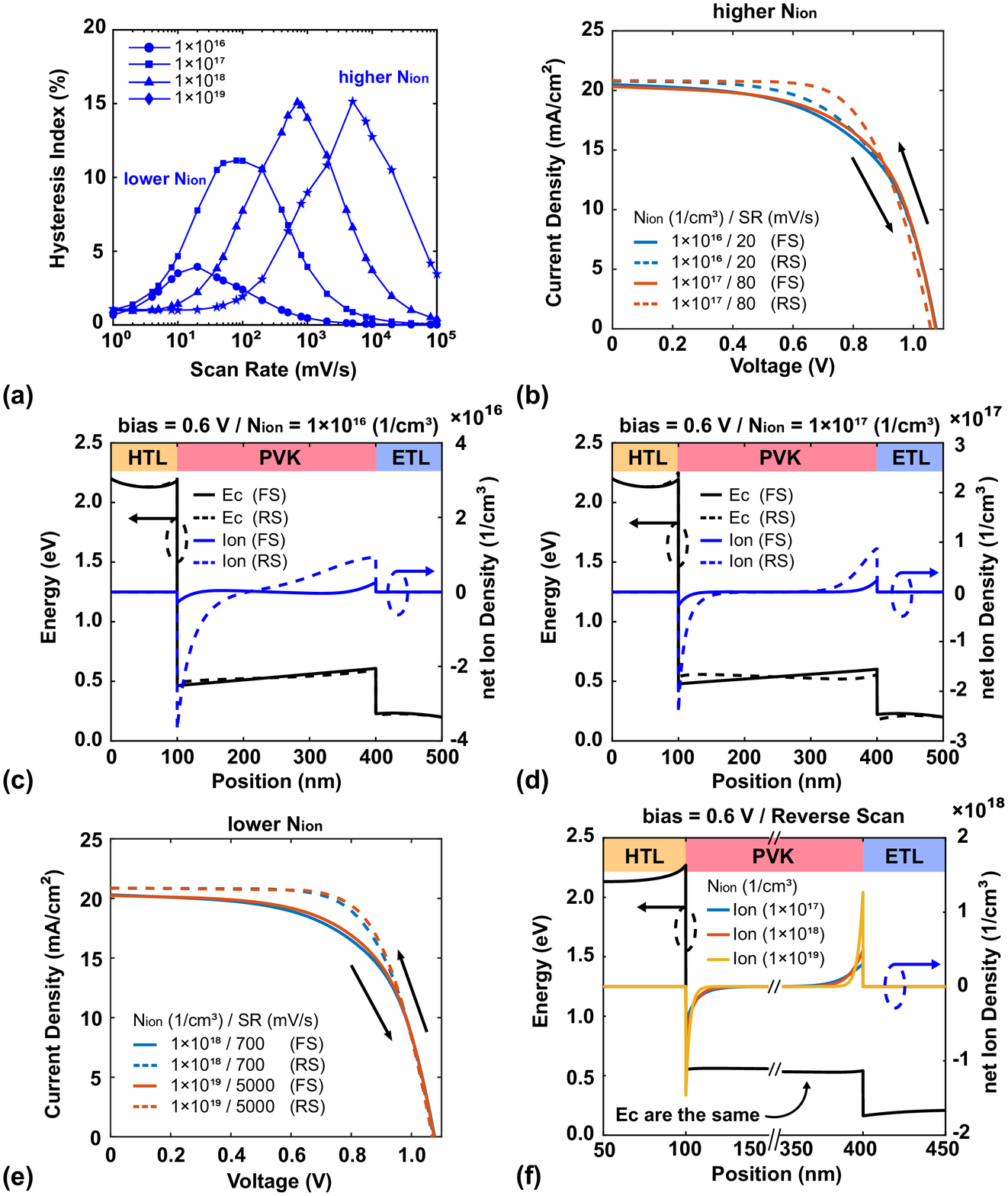}}
\caption{(a) HI vs. scan rate for $N_{ion}$ = $1\times 10^{16}$, $1\times 10^{17}$, $1\times 10^{18}$ and $1\times 10^{19}$ $\rm cm^{-3}$. (b) $J$--$V$ characteristics for the lower ion density. (c)--(d) Conduction band and the net ion density distribution at 0.6 V for the forward and reverse scanning in the cases of $N_{ion}$ = $1\times 10^{16}$ and $1\times 10^{17}$ $\rm cm^{-3}$, respectively. (e) $J$--$V$ characteristics for the higher ion density. (f) Conduction band and ion accumulation of the ETL/PVK and PVK/HTL interfaces at 0.6 V for $N_{ion}$ = $1\times 10^{17}$, $1\times 10^{18}$ and $1\times 10^{19}$ $\rm cm^{-3}$. }
\label{Fig6}
\end{figure}

Figure \ref{Fig6}(a) shows the changes in HI with scan rate for different $N_{ion}$. The results can be divided into two behaviors: (1) $N_{ion}$ = $\rm 10^{16}$ to $\rm 10^{17}~cm^{-3}$); (2) $N_{ion}$ = $\rm 10^{18}$ to $\rm 10^{19}~cm^{-3}$). For the lower total ion density [$N_{ion}$ of $\rm 10^{16}$ and $\rm 10^{17}$ in figure \ref{Fig6}(a)], the hysteresis gradually increases as the total ion density increases, as shown in figure \ref{Fig6}(b). The hysteresis is quite weak when $N_{ion}$ is only $\rm 10^{16}~cm^{-3}$. Figure \ref{Fig6}(c) shows that the difference in the conduction band of the perovskite layer (black line) is not obvious between the forward and reverse scanning at the same voltage of 0.6 V. Compared to the other cases, the ion accumulation (blue line in figure \ref{Fig6}(c)) is relatively low in this case ($\rm \approx10^{16}~cm^{-3}$). Because the total ion density is not sufficient to screen the built-in electric field, hysteresis is quite weak. When $N_{ion}$ is increased to $\rm 10^{17}~cm^{-3}$, the hysteresis becomes obvious. Figure \ref{Fig6}(d) shows that the change in the conduction band of the perovskite layer (black line) is greater than in the case of lower total ion density, while the ion accumulation (blue line) is higher in this case ($\rm \approx10^{17}~cm^{-3}$). Because the extra ion density results in a greater degree of screening of the built-in electric field compared to in the case of lower total ion density, the hysteresis is obvious. 

For the higher total ion density [$N_{ion}$ of $\rm 10^{18}$ and $\rm 10^{19}$ in figure \ref{Fig6}(a)], the patterns in HI are similar at the different scan rates. The system response region shifts with total ion density and the point of maximum HI changes; the maximum HI values are 700 and 5000~mV/s for $N_{ion}$ = $\rm 1\times 10^{18}$ and $\rm 10^{19}$, respectively. As shown in figure \ref{Fig6}(e), the $J$--$V$ characteristics are the same for the two cases; thus, the HI values are also the same. However, figure \ref{Fig6}(f) shows that the net ion density distribution is not the same for the three cases. The net ion density at the interface is higher when the total ion density is higher. Although the distributions are different, the summations are almost the same, indicating that the ion-induced electric fields are the same. This phenomenon can be explained by the different timescales of ion accumulation. Because the net ion distribution is decided by the ion drift-diffusion equation [Eq. (\ref{eq15})], the time required to achieve sufficient ion accumulation to screen the built-in electric field is shorter than in the other cases when the total ion density is quite high. Since the time is shorter and the voltage is the same, the system response scan rate is higher than in the other cases. In other words, it takes less time to accumulate sufficient ions to screen the built-in electric field.

\begin{figure}[!bpht]
\renewcommand{\figurename}{Figure}
\centering
\resizebox{85mm}{!}{\includegraphics{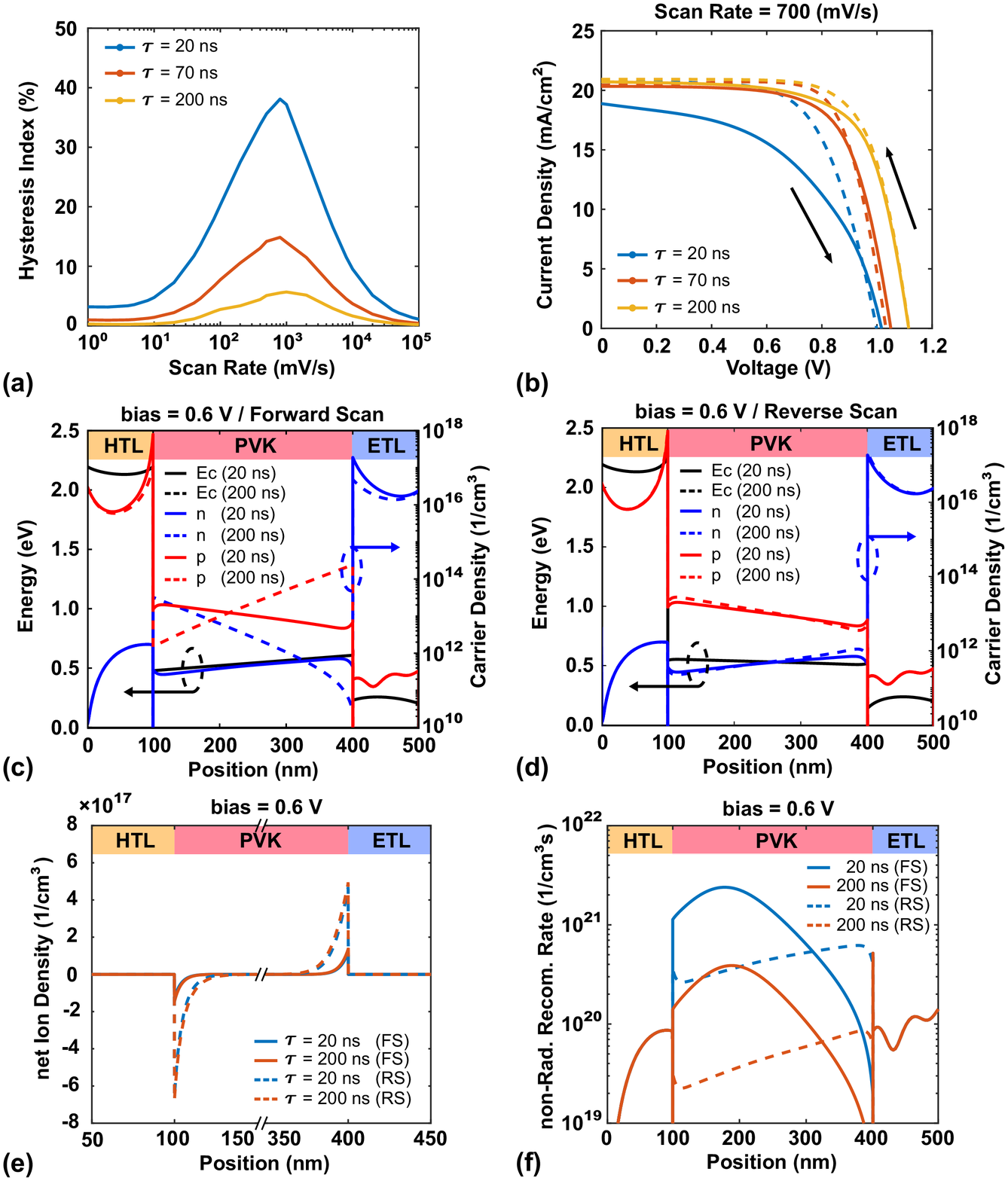}}
\caption{(a) HI vs. scan rate for carrier lifetimes of 1, 5, and 10~ns. (b) $J$--$V$ characteristics for carrier lifetimes of 20, 70, and 200~ns. (c)--(d) Conduction band and carrier density for carrier lifetimes of 20 and 200~ns at 0.6 V for the forward and reverse scanning, respectively. (e)--(f) Net ion density distribution and non-radiative recombination rate, respectively, for the two cases at a voltage of 0.6 V for the forward and reverse scanning.}
\label{Fig7}
\end{figure}

Figures \ref{Fig7}(a) and \ref{Fig7}(b) show that hysteresis is more serious at the carrier lifetime of 20 ns than in cases with longer carrier lifetimes. To conveniently examine this situation, we chose the scan rate with the maximum HI (700$\rm ~mV/s$). Some studies have attributed strong hysteresis in PSC devices to high non-radiative recombination, and poor crystalline quality \cite{kim2014parameters}. Indeed, the quality of perovskite can affect the performance of the PSC device; however, the reason for the high degree of hysteresis is less well studied. Figures \ref{Fig7}(c) and \ref{Fig7}(d) show the electron/hole distributions and conduction bands for carrier lifetimes of 20 and 200 ns under forward and reverse scanning, respectively. Under the forward scanning, the carrier distribution of both electrons and holes is less for the carrier lifetime of 20 ns than for the lifetime of 200 ns. Furthermore, at the voltage of 0.6 V, the conduction bands are similar for the forward and reverse scanning. Figure \ref{Fig7}(e) shows that the net ion distribution is almost the same; thus, the difference in the degree of hysteresis has nothing to do with ion accumulation. Figure \ref{Fig7}(f) shows that the non-radiative recombination is stronger for the carrier lifetime of 20 ns than for the lifetime of 200 ns. This is the primary reason that perovskite quality affects the performance of the PSC device. The different degrees of hysteresis under different carrier lifetimes are related to the proportions of non-radiative recombination. The non-radiative recombination of the active layer for each case is shown in Table \ref{tab2}. While ion accumulation can improve the performance under reverse scanning conditions in both cases, the performance in the case of a 20 ns carrier lifetime is quite low under forward scanning conditions. The improvement in non-radiative recombination from ion accumulation is relatively greater than the case of 200 ns. Hence, the improvement of performance is greater than the case of 200 ns, and it causes a stronger hysteresis.

\begin{table}[t]

\caption{Non-radiative recombination rates at 0.6 V}
\centering
\begin{tabular}{cccc}
 \hline\hline\\
  Case         & nRad.$\rm ^{FS}$ &  nRad.$\rm ^{RS}$ &  nRad.$\rm ^{(FS-RS)}$ / nRad.$\rm ^{FS}$ \\ 
    &  (cm$\rm ^{-2}$s$\rm ^{-1}$) & (cm$\rm ^{-2}$s$\rm ^{-1}$)& (\%)
\\\hline\\
  1 ns & $1.74\rm \times 10^{18}$ & $8.92\rm \times 10^{17}$ & 48.7 \\\\
  
  10 ns & $3.59\rm \times 10^{17}$ & $2.19\rm \times 10^{17}$ & 39.0 \\\\
\hline \hline

\end{tabular}
\label{tab2}
\end{table}

\section{Conclusion}
In this work,  perovskite solar cells (PSCs) with different ETLs were fabricated to understand the hysteresis phenomenon under a series of scan rates. The experimental results show that hysteresis phenomenon would be affected by the dielectric constant of ETL and scan rate significantly. Also, a modified Poisson-DD solver coupled with a fully time-dependent ion migration model was used to analyze ion migration in different situations and reveal the mechanisms of the observed effects. Hysteresis was shown to be caused by ion accumulation. Due to the built-in electric field of perovskite, ions accumulate at both sides of the interface and form an ion-induced electric field. Therefore, the $J$--$V$ curves are not the same under the forward and reverse scanning. Generally, the performance of the PSCs device is greater under the reverse scanning than under the forward scanning because carrier extraction is enhanced by the ion-induced electric field under reverse scanning, while it decreased under forward scanning. This is the primary factor that leads to hysteresis. The change in HI with scan rate was demonstrated, and the curve of HI vs. scan rate was shown to be Gaussian-shaped rather than monotonic. Hysteresis is weak when the scan rate is lower or higher than the system response scan rate. At low scan rates, the net ion distribution depends on the bias. At fast scan rates, the ions cannot respond to the voltage scan, and the ion effect is thus quite weak. Hence, the HI is low under both slow and fast scan rates, although the mechanisms are different. Hysteresis only occurs when the scan rate lies in the system response region. Possible reasons leading to different degrees of hysteresis in previous works include the quality of the perovskite, degree of perovskite degradation, and properties of the ETL; these factors correspond to the carrier lifetime of perovskite, total ion density, and dielectric constant, respectively. The mechanisms of the effects of the above-mentioned factors were demonstrated and explained. The effect of total ion density is explained in a relatively intuitive fashion. Hysteresis is weak when the total ion density is low; thus, doping with K, Rb, and Li can significantly reduce the hysteresis. The effect of carrier lifetime is related to the different proportions of non-radiative recombination under the forward and reverse scanning. Hence, hysteresis is stronger when the lifetime of perovskite is lower (i.e., the perovskite quality is poor). Finally, the properties of the ETL, especially the dielectric constant, play an important role in hysteresis in PSC devices. According to the modeling results, using an ETL with a low dielectric constant can reduce the degree of hysteresis. This is because the electric field of the ETL increases as the dielectric constant of the ETL decreases; thus, decreasing the dielectric constant can reduce the built-in electric field of perovskite and hence reduce hysteresis.

\begin{acknowledgments}
The authors gratefully acknowledge the Ministry of Science and Technology (MOST) of Taiwan under Grant Nos. 108-2628-E002-010-MY3, 108-2627-H-028-002, and 109-2221-E-002-196-MY2.
\end{acknowledgments}

\section*{Supporting Information}
The supporting material is available, as well as the animations of ion migration are shown on the website. The modeling tool was developed by our laboratory. More detailed information about the modeling tool can be found on our website (http://yrwu-wk.ee.ntu.edu.tw/).

\nocite{*}
\bibliography{aipsamp}

\end{document}